\documentclass{cs19proc}

\usepackage{kantlipsum}
\usepackage{graphicx}
\usepackage{subfig}
\usepackage{epstopdf}
\usepackage{verbatim}
\usepackage{multirow}
\usepackage{float}

\usepackage{natbib}

\def\spose#1{\hbox to 0pt{#1\hss}}
\def\lta{\mathrel{\spose{\lower 3pt\hbox{$\mathchar"218$}}
     \raise 2.0pt\hbox{$\mathchar"13C$}}}
\def\gta{\mathrel{\spose{\lower 3pt\hbox{$\mathchar"218$}}
     \raise 2.0pt\hbox{$\mathchar"13E$}}}

\editors{G.~A. Feiden}
\publisher{Zenodo}
\conference{The 19th Cambridge Workshop on Cool Stars, Stellar Systems, and the Sun}
\conferencedate{2016}

\title{Fitting Formulas for Determining the Existence of S-type and P-type \\
Habitable Zones in Binary Systems: First Results}
\author{Zhaopeng Wang$^{1}$, Manfred Cuntz$^{1}$}

\affiliation{$^{1}$ Department of Physics, University of Texas at Arlington, Arlington, USA}

\shorttitle{Fitting Formulas for S-type and P-type Habitable Zones}
\shortauthors{Zhaopeng Wang \& Manfred Cuntz}

\abs{We present initial work about attaining fitting formulas for the quick determination of the
existence of S-type and P-type habitable zones in binary systems.  Following previous work, we
calculate the limits of the climatological habitable zone in binary systems (which sensitively
depend on the system parameters) based on a joint constraint encompassing planetary orbital
stability and a habitable region for a possible system planet.  We also consider updated results
on planetary climate models previously obtained by Kopparapu and collaborators.  Fitting equations
based on our work are presented for selected cases.}

\begin{document}

\maketitle

\section{Introduction}
Since the first confirmed detection of exoplanets in 1992, more than 3500 exoplanets have been
found, with the 1000th detection by the $Kepler$ mission
announced\footnote{{\tt http://science.nasa.gov/science-news/science-at-nasa\\/2015/06jan\_kepler1000}.}
by NASA on January 6th, 2015.
With the number of discoveries sky-rocketing, it is adequate to continue studying
exoplanets, as well as searching for exomoons, besides identifying habitable zones (HZs).

It is also well-known that binary (as well as higher order) systems occur
frequently in the local Galactic neighborhood (e.g., Duquennoy \& Mayor 1991,
and subsequent studies).  Observations show that exoplanets can also exist in
binary systems, and might also be orbitally stable for millions or billions of years.
There are two types of possible orbits (e.g., Dvorak 1982): planets orbiting one
of the binary components are said to be in S-type orbits, while planets orbiting both
binary components are said to be in P-type orbits.  For example, Kepler-413~b (Kostov et al. 2014)
is in a P-type orbit, indicating that the planet is orbiting both stellar components of the
binary system.  Kepler-453~b (Welsh et al. 2015) also constitutes a transiting circumbinary planet.
Planetary S-type orbits have more confirmed detections, such as Kepler-432~b (Oritz et al. 2015).
Some of these planets are located within the stellar HZs, as, for example, Kepler-62~f (Borucki
et al. 2013); those cases typically receive significant attention due to their potential of
hosting alien life.

In previous studies, focusing on habitable zones in stellar binary systems, presented
by Cuntz (2014, 2015), denoted as Paper I and II, respectively, henceforth, a joint constraint
of radiative habitable zones (RHZs, based on stellar radiation) and orbital stability was
considered.  Moreover, Paper~II also takes into account the eccentricity of binary components.
RHZs, including conservative,
general and extended habitable zones (therefore referred to as CHZ, GHZ, and EHZ, respectively),
are defined in the same way as for the solar system (see Section 2.1).

Our paper is structured as follows. In Section~2, we briefly describe the theoretical
approach for the calculation of HZs adopted from Paper I and II; however, our work also
takes into account revised HZ limits for the Solar System from updated climate models.
In Section~3, we present some case studies with fitting equations for identifying the
existence of HZs. Our summary will be given in Section~4.

\section{Theoretical Approach}
\subsection{Habitability limits}
In Paper I and II, habitable zones have been defined based on habitability limits for
the Solar System from previous studies by Kasting et al. (1993) [Kas93] and Mischna et al. (2000).
Thereafter, Kopparapu et al. (2013, 2014) [Kop1314] presented updated results on habitability limits
by introducing new climate models.  Table~1 conveys the meaning of each habitability limit.
In this paper, we consider results for two types of RHZs: GHZ and RVEM. The GHZ is the region between the
habitability limits of runaway greenhouse effect and maximum greenhouse effect (without
clouds). The RVEM, explicitly, has the recent Venus position as inner limit and the early Mars
position as outer limit.

\begin{table*}[t]
	\centering
	\caption{Habitability limits for the Solar System}
	\label{tab:table_wide}
	\begin{tabular}[t]{ l @{\extracolsep{\fill}} c c c c l}
		\noalign{\smallskip}\hline\hline\noalign{\smallskip}
		Description & Indices & \multicolumn{3}{ c }{Models} & This work\\
		\noalign{\smallskip}\hline\noalign{\smallskip}
		... & l    & \multicolumn{2}{ c }{Kas93} & Kop1314 & ... \\
		\noalign{\smallskip}\hline\noalign{\smallskip}
		...  & ... & 5700~K & 5780~K & 5780~K & ... \\
		...  & ... & (AU)   & (AU)   & (AU)   & ... \\
		\noalign{\smallskip}\hline\noalign{\smallskip}
		Recent Venus					&  1  &  0.75    &  0.77		& 0.750	& RVEM Inner Limit \\
		Runaway greenhouse effect			&  2  &  0.84    &  0.86		& 0.950	& GHZ Inner Limit \\
		Moist greenhouse effect				&  3  &  0.95    &  0.97		& 0.993	& ... \\
		Earth-equivalent position			&  0  &  0.993   &  $\equiv$1 	& $\equiv$1	& ... \\
		First CO$_2$ condensation			&  4  &  1.37    &  1.40		& ... 	& ... \\
		Maximum greenhouse effect, no clouds	&  5  &  1.67    &  1.71		& 1.676	& GHZ Outer Limit \\
		Early Mars						&  6  &  1.77    &  1.81		& 1.768	& RVEM Outer Limit \\
		\noalign{\smallskip}\hline\noalign{\smallskip}
	\end{tabular}
\end{table*}

\subsection{Calculation of the HZs}

In order to achieve the same condition as in Solar System on behave of radiation, the planet should
receive the same amount of radiative energy fluxes from both binary stellar components in total.
Thus the equation for calculating RHZs (see Paper~I) yields
\begin{equation}
	\setlength{\abovedisplayskip}{5pt}
	\setlength{\belowdisplayskip}{5pt}
	\frac{L_{1}}{S_{{\rm rel},1l}d^{2}_{1}} +  \frac{L_{2}}{S_{{\rm rel},2l}d^{2}_{2}}= \frac{L_{\odot}}{s^{2}_{l}}
\end{equation}
with $L_{1}$ and $L_{2}$ denoting the stellar luminosities, $d_{1}$ and $d_{2}$ denoting the distances 
to the binary components (see Figure 1) and $s_{l}$
being one of the solar habitability limits (see Table 1). $S_{\rm rel}$, as a function of effective temperature
of binary stars, represents stellar flux in units of solar constant. Because $d_{1}$ and $d_{2}$
can be converted into a function of $z$, the distance from the center of binary system, a
quartic equation for $z$ can be obtained after algebraic transformations (see Paper I and II for details). 

\begin{figure}[h]
	\centering
	\subfloat[S-type]{{\includegraphics[width=0.45\linewidth]{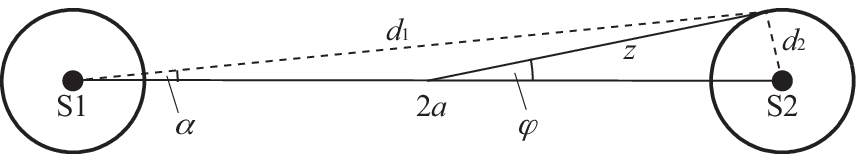} }}\\
	\subfloat[P-type]{{\includegraphics[width=0.45\linewidth]{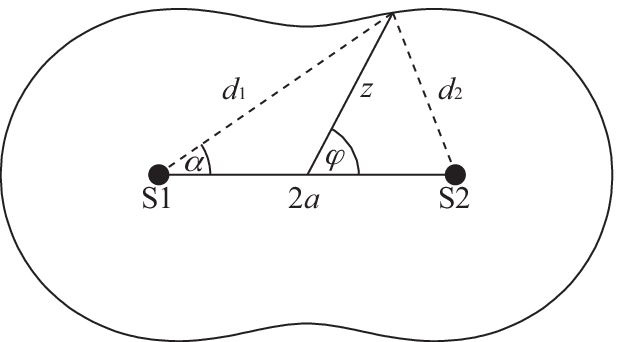} }}
	\caption{Mathematical models of S-type (top) and P-type (bottom) habitable zones for
binary systems. It is not required to have the stars S1 and S2 being identical in this method
(adopted from Paper I and II).}
	\label{fig:fig_wide}
\end{figure}

The RHZ, as an annulus around each star (S-type) or both stars (P-type), is therefore given as
\begin{equation}
	\setlength{\abovedisplayskip}{5pt}
	\setlength{\belowdisplayskip}{5pt}
	\textrm{RHZ(z)} = \textrm{Min}({\cal R}(z,\varphi))|s_{l,{\rm out}} - \textrm{Max}({\cal R}(z,\varphi))|s_{l,{\rm in}}
\end{equation}

If a planet is assumed to stay in the HZ for a long period of time, it is mandated that it has
a stable orbit. The planetary orbital stability limit is an upper limit
as the distance from the main star for S-type orbit. For P-type orbit, it becomes a lower limit
which is measured from the center of the binary system. The orbital stability limits
are obtained from the fitting equations developed by Holman \& Wiegert (1999); see Paper II
for details.  Moreover, according to the terminology of Paper~I, ST-type and PT-type HZs denote
the cases when the HZs are smaller than the corresponding RHZs due to truncation owing to the
orbital stability limits, while S-type and P-type HZs refer to the cases where the HZs coincide
with the RHZs.

\begin{figure}[H]
	\centering
	\includegraphics[width=0.85\linewidth]{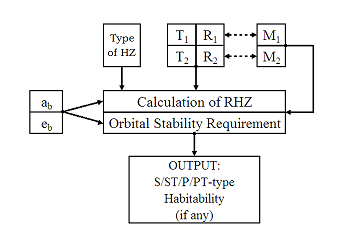}
	\caption{Flow diagram of the calculation (akin to Paper II).
      See also Cuntz \& Bruntz (2014) for information on the webpage BinHab, hosted by
      the University of Texas at Arlington, allowing the calculation of stellar HZs.}
	\label{fig:fig_narrow}
\end{figure}

\section{Case Study}
\subsection{Existence of HZs}
Following the method as dicussed, various sets of systems, including systems of equal
and non-equal masses, have been studied to examine their HZs.  The stellar parameters
can be found in Table 2.

\begin{table}[H]
	\centering
	\caption{Stellar Parameters}
	\label{tab:table_narrow}
	\begin{tabular*}{0.85\linewidth}{l @{\extracolsep{\fill}} c l}
		\noalign{\smallskip}\hline\hline\noalign{\smallskip}
		$M_{*}$ & Spectral Type & $L_{*}$ \\
		($M_{\odot}$) & ... & ($L_{\odot}$) \\
		\noalign{\smallskip}\hline\noalign{\smallskip}
		1.25 & F7V & 2.1534    \\
		1.00 & G2V & 1.0000    \\
		0.75 & K2V & 0.35569   \\
		0.50 & M0V & 0.043478  \\
		\noalign{\smallskip}\hline
	\end{tabular*}
\end{table}

Figure 3 shows the requirements for the GHZ and RVEM to exist for selected binary systems,
i.e., systems with masses $M_{1} = M_{2} =  0.50~M_{\odot}$; $M_{1} = 1.00~M_{\odot}$,
$M_{2} = 0.50~M_{\odot}$; $M_{1} = M_{2} = 1.00~M_{\odot}$, and $M_{1} = 1.25~M _{\odot}$,
$M_{2} = 0.50~M_{\odot}$.

For systems with masses $M_{1} = M_{2} = 0.50~M_{\odot}$, in case of $e_{b}$ = 0,
$2a$ is required to be smaller than 0.44 AU for the P/PT-type GHZ to exist and smaller than 0.48~AU for
the P/PT-type RVEM to exist.  Regarding S/ST-type HZs, $2a$ needs to be larger than 1.59 AU
and larger than 1.25 AU for the GHZ and RVEM to exist, respectively.

Keeping $e_{b}$ = 0, the other equal-mass system considered in our study with
$M_{1} = M_{2} = 1.00~M_{\odot}$,
requires $2a$ to be smaller than 2.03 AU and smaller than 2.14 AU
for the GHZ and RVEM to exist, respectively, regarding P/PT-type HZs.
For S/ST-type cases, $2a$ needs to be larger than 7.76 AU and larger than 6.12 AU,
respectively, for the GHZ and RVEM to exist.

Non-equal mass cases, which generally are more significant, have also been investigated
in this study.
In systems with $M_{1} = 1.00~M_{\odot}, M_{2} = 0.50~M_{\odot}$, $2a$ is required
to be smaller than 0.95 AU and smaller than 1.29 AU for the P/PT-type GHZ and RVEM,
respectively, in case of $e_{b}$ = 0. Furthermore, $2a$ needs to be larger than 6.45 AU
and larger than 5.09 AU for the S/ST-type GHZ and RVEM, respectively.

Moreover, we also considered systems with $M_{1} = 1.25~M _{\odot}$ and
$M_{2} = 0.50~M_{\odot}$.  The case of $e_{b}$ = 0 requires $2a$ to be smaller
than 1.12 AU and smaller than 1.56 AU for the GHZ and RVEM, respectively,
in case of P/PT-type HZs.  Regarding S/ST-type HZs, $2a$ is required to be larger
than 7.96 AU for GHZ and 6.24 AU for RVEM.

If larger eccentricities are considered, the sizes of the P/PT HZs are barely affected,
whereas the sizes of the S/ST HZs are significantly reduced.  This behavior is found for
all four stellar mass combinations and for both types of HZs.

\subsection{Initial work on fitting equations for determinating the existence of HZs}

In the previous section, we present the requirements for HZs to exist (see Figure 3).
The curves consist of pairs of critical values of $2a$ and $e_{b}$ and form regions for each
figure that indicate the existence of the corresponding HZ. In this section, we present
the fitting equations of these $2a$ versus $e_{b}$ curves allowing the efficient determination
of the existence of each HZ.
Linear least square method has been used for the development of each fitting equation.
Moreover, the coefficient of determination ($R^{2}$) is used to check the goodness of fitting.

For P/PT-type cases, the equation reads
\begin{equation}
	\setlength{\abovedisplayskip}{5pt}
	\setlength{\belowdisplayskip}{5pt}
	2a = \alpha_{1} + \alpha_{2}e_{b} + \alpha_{3}e_{b}^{2}
\end{equation}

The equation for S/ST-type yields
\begin{equation}
	\setlength{\abovedisplayskip}{5pt}
	\setlength{\belowdisplayskip}{5pt}
	2a = e^{\beta_{1} + \beta_{2}e_{b} + \beta_{3}e_{b}^{2}}
\end{equation}

The coefficients of systems discussed in the previous section are given in Table 3--6,
as well as coefficients of determination. Fitting results are plotted and compared
with the data in Figure 4. 

\begin{table}[H]
	\centering
	\caption{$M_{1} = M_{2} = 0.50~M_{\odot}$}
	\label{tab:table_narrow}
	\begin{tabular*}{0.85\linewidth}{l @{\extracolsep{\fill}} c c c c}
		\noalign{\smallskip}\hline\hline\noalign{\smallskip}
		HZ & $\alpha_{1}$ & $\alpha_{2}$ & $\alpha_{3}$ & $R^{2}$ \\
		\noalign{\smallskip}\hline\noalign{\smallskip}
		P-GHZ & 0.44 & -0.44 & 0.31 & 0.9975 \\
		P-RVEM & 0.47 & -0.47 & 0.31 & 0.9994 \\
		\noalign{\smallskip}\hline\hline\noalign{\smallskip}
		HZ & $\beta_{1}$ & $\beta_{2}$ & $\beta_{3}$ & $R^{2}$ \\
		\noalign{\smallskip}\hline\noalign{\smallskip}
		S-GHZ & 0.61 & -0.14 & 2.97 &  0.9949 \\
		S-RVEM & 0.40 & -0.23 & 3.06 & 0.9953 \\
		\noalign{\smallskip}\hline
	\end{tabular*}
\end{table}

\begin{table}[H]
	\centering
	\caption{$M_{1} = M_{2} = 1.00~M_{\odot}$}
	\label{tab:table_narrow}
	\begin{tabular*}{0.85\linewidth}{l @{\extracolsep{\fill}} c c c c}
		\noalign{\smallskip}\hline\hline\noalign{\smallskip}
		HZ & $\alpha_{1}$ & $\alpha_{2}$ & $\alpha_{3}$ & $R^{2}$ \\
		\noalign{\smallskip}\hline\noalign{\smallskip}
		P-GHZ & 2.00 & -1.94 & 1.19 & 0.9984 \\
		P-RVEM & 2.11 & -2.00 & 1.18 & 0.9982 \\
		\noalign{\smallskip}\hline\hline\noalign{\smallskip}
		HZ & $\beta_{1}$ & $\beta_{2}$ & $\beta_{3}$ & $R^{2}$ \\
		\noalign{\smallskip}\hline\noalign{\smallskip}
		S-GHZ & 2.06 & 1.02 & 1.28 &  0.9998 \\
		S-RVEM & 1.84 & 0.87 & 1.54 & 0.9995 \\
		\noalign{\smallskip}\hline
	\end{tabular*}
\end{table}

\begin{table}[H]
	\centering
	\caption{$M_{1} = 1.00~M_{\odot}, M_{2} = 0.50~M_{\odot}$}
	\label{tab:table_narrow}
	\begin{tabular*}{0.85\linewidth}{l @{\extracolsep{\fill}} c c c c}
		\noalign{\smallskip}\hline\hline\noalign{\smallskip}
		HZ & $\alpha_{1}$ & $\alpha_{2}$ & $\alpha_{3}$ & $R^{2}$ \\
		\noalign{\smallskip}\hline\noalign{\smallskip}
		P-GHZ & 0.95 & -0.84 & 0.41 & 0.9992 \\
		P-RVEM & 1.27 & -1.34 & 0.88 & 0.9974 \\
		\noalign{\smallskip}\hline\hline\noalign{\smallskip}
		HZ & $\beta_{1}$ & $\beta_{2}$ & $\beta_{3}$ & $R^{2}$ \\
		\noalign{\smallskip}\hline\noalign{\smallskip}
		S-GHZ & 1.88 & 0.98 & 1.46 &  0.9995 \\
		S-RVEM & 1.66 & 0.82 & 1.74 & 0.9992 \\
		\noalign{\smallskip}\hline
	\end{tabular*}
\end{table}

\begin{table}[H]
	\centering
	\caption{$M_{1} = 1.25~M_{\odot}, M_{2} = 0.50~M_{\odot}$}
	\label{tab:table_narrow}
	\begin{tabular*}{0.85\linewidth}{l @{\extracolsep{\fill}} c c c c}
		\noalign{\smallskip}\hline\hline\noalign{\smallskip}
		HZ & $\alpha_{1}$ & $\alpha_{2}$ & $\alpha_{3}$ & $R^{2}$ \\
		\noalign{\smallskip}\hline\noalign{\smallskip}
		P-GHZ & 1.11 & -0.98 & 0.47 & 0.9987 \\
		P-RVEM & 1.56 & -1.68 & 1.09 & 0.9971 \\
		\noalign{\smallskip}\hline\hline\noalign{\smallskip}
		HZ & $\beta_{1}$ & $\beta_{2}$ & $\beta_{3}$ & $R^{2}$ \\
		\noalign{\smallskip}\hline\noalign{\smallskip}
		S-GHZ & 2.08 & 1.11 & 1.23 &  0.9997 \\
		S-RVEM & 1.85 & 0.99 & 1.48 & 0.9996 \\
		\noalign{\smallskip}\hline
	\end{tabular*}
\end{table}

As all the coefficient of determination are close to 1, the fitting results
should thus be very close to the data. In Figure 4, fitting results plotted
as well as the data for comparison. Some of them are virtually indistinguishable
from the data. Table 7 shows the precent errors of a few selected data points for
$M_{1} = M_{2} =  1.00~M_{\odot}$ and $M_{1} = 1.00~M_{\odot}, M_{2} =
0.50~M_{\odot}$ cases.
\begin{equation}
	\setlength{\abovedisplayskip}{5pt}
	\setlength{\belowdisplayskip}{5pt}
	\textrm{Percentage Error} = \left| \frac{data - fitting}{data} \right|
\end{equation}

\begin{table*}[t]
	\centering
	\caption{Results for two stellar binary systems}
	\label{tab:table_wide}
	\begin{tabular*}{0.85\linewidth}{c || @{\extracolsep{\fill}} c c c c | c c c c}
		\noalign{\smallskip}\hline\hline\noalign{\smallskip}
		$e_{b}$ & \multicolumn{4}{ c }{$M_{1} = M_{2} = 1.00~M_{\odot}$} & \multicolumn{4}{ c }{$M_{1} = 1.00~M_{\odot}, M_{2} = 0.50~M_{\odot}$} \\
		\noalign{\smallskip}\hline\noalign{\smallskip}
		... & P-GHZ & P-RVEM & S-GHZ & S-RVEM & P-GHZ & P-RVEM & S-GHZ & S-RVEM \\
		\noalign{\smallskip}\hline\noalign{\smallskip}
		0.0 & 1.35\% & 1.33\% & 1.11\% & 2.96\% & 0.21\% & 1.51\% & 1.60\% & 3.34\% \\ 
		0.1 & 0.37\% & 0.25\% & 0.27\% & 0.09\% & 0.56\% & 0.59\% & 0.49\% & 0.14\% \\ 
		0.2 & 0.58\% & 0.67\% & 0.55\% & 0.87\% & 0.64\% & 0.99\% & 1.15\% & 1.58\% \\ 
		0.3 & 0.09\% & 0.28\% & 0.16\% & 0.64\% & 0.40\% & 0.40\% & 0.90\% & 1.51\% \\ 
		0.4 & 0.60\% & 0.46\% & 0.38\% & 0.08\% & 0.01\% & 0.50\% & 0.01\% & 0.30\% \\ 
		0.5 & 1.01\% & 0.99\% & 0.04\% & 0.68\% & 0.31\% & 1.08\% & 0.23\% & 0.77\% \\ 
		0.6 & 0.83\% & 1.03\% & ...    & 0.94\% & 0.19\% & 0.91\% & ...    & 0.27\% \\ 
		0.7 & 0.42\% & 0.21\% & ...    & ...    & 0.57\% & 0.61\% & ...    & ...    \\ 
		0.8 & 2.89\% & 1.79\% & ...    & ...    & 2.21\% & 4.16\% & ...    & ...    \\ 
		\noalign{\smallskip}\hline
	\end{tabular*}
\end{table*}

\begin{figure*}[t]
	\centering
	\subfloat{{\includegraphics[width=0.45\linewidth]{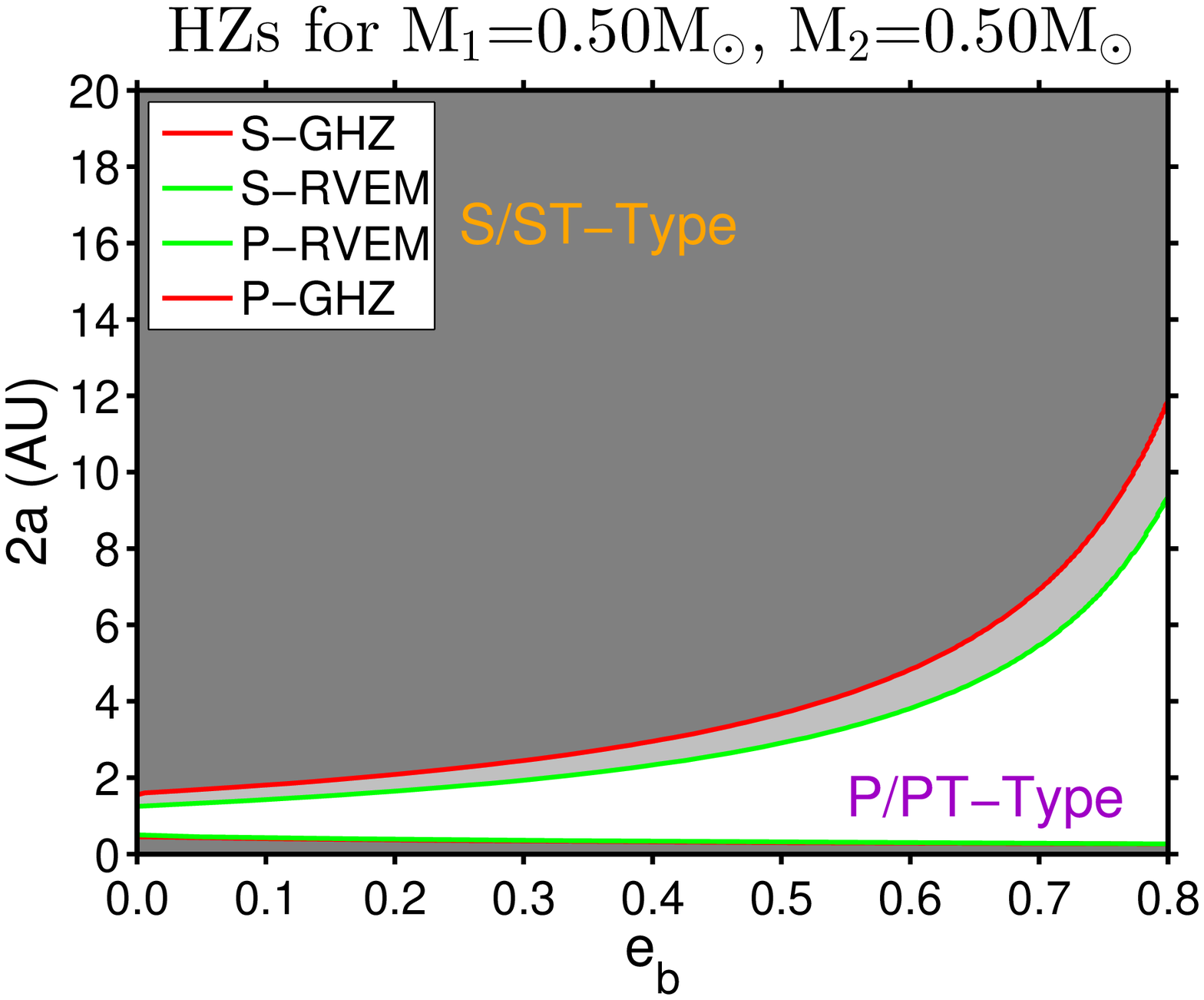} }}
	\subfloat{{\includegraphics[width=0.45\linewidth]{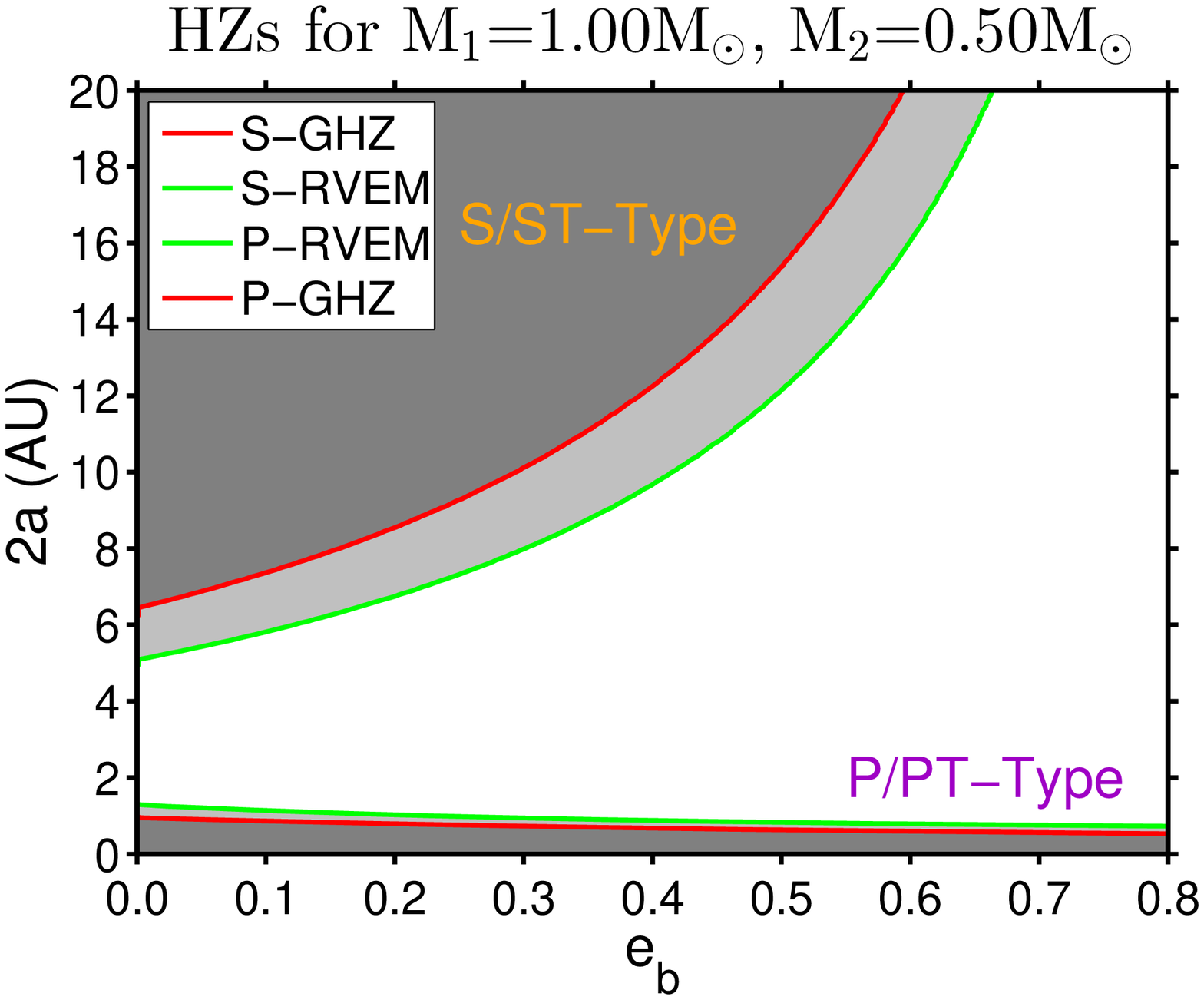} }}\\
	\subfloat{{\includegraphics[width=0.45\linewidth]{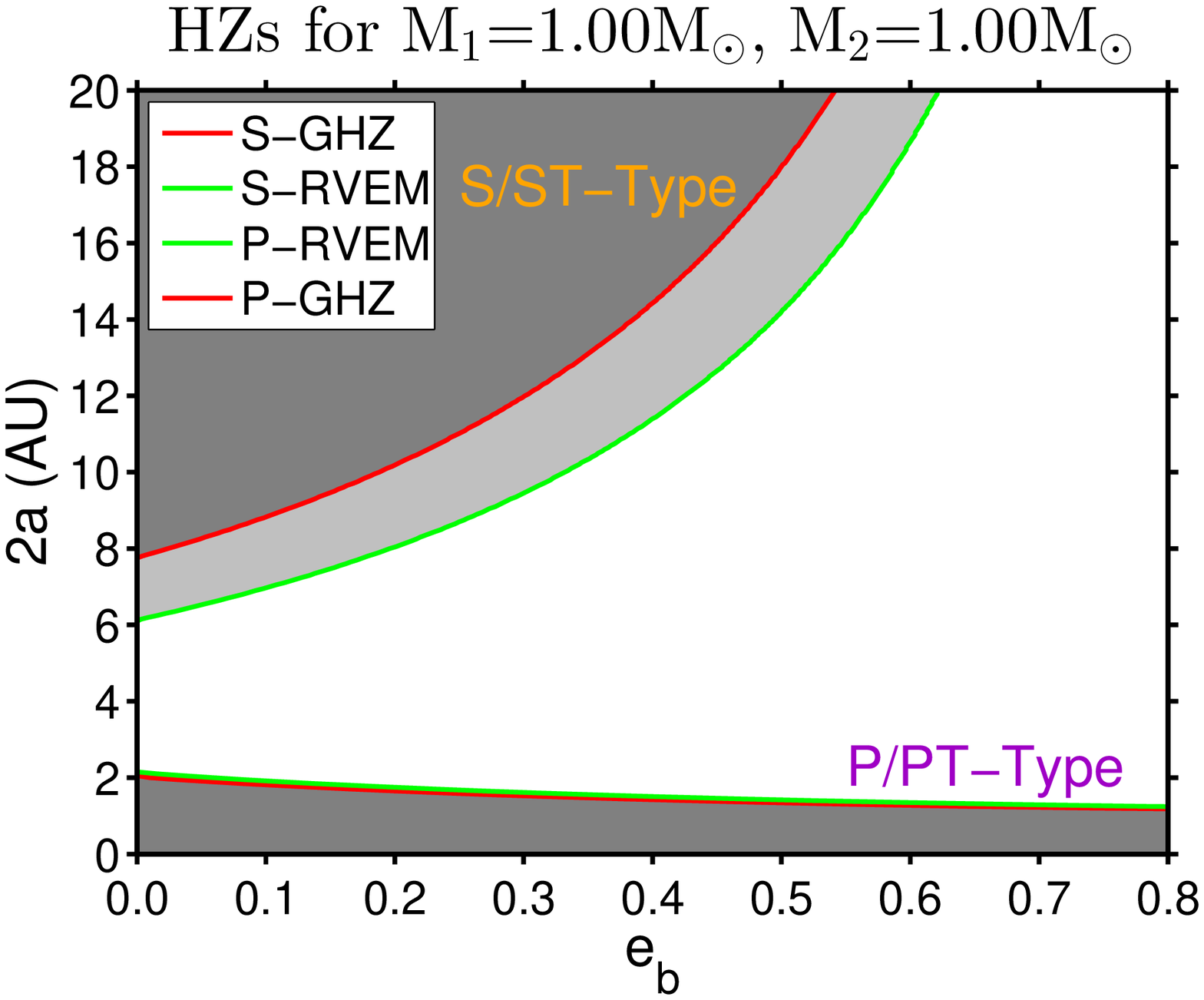} }}
	\subfloat{{\includegraphics[width=0.45\linewidth]{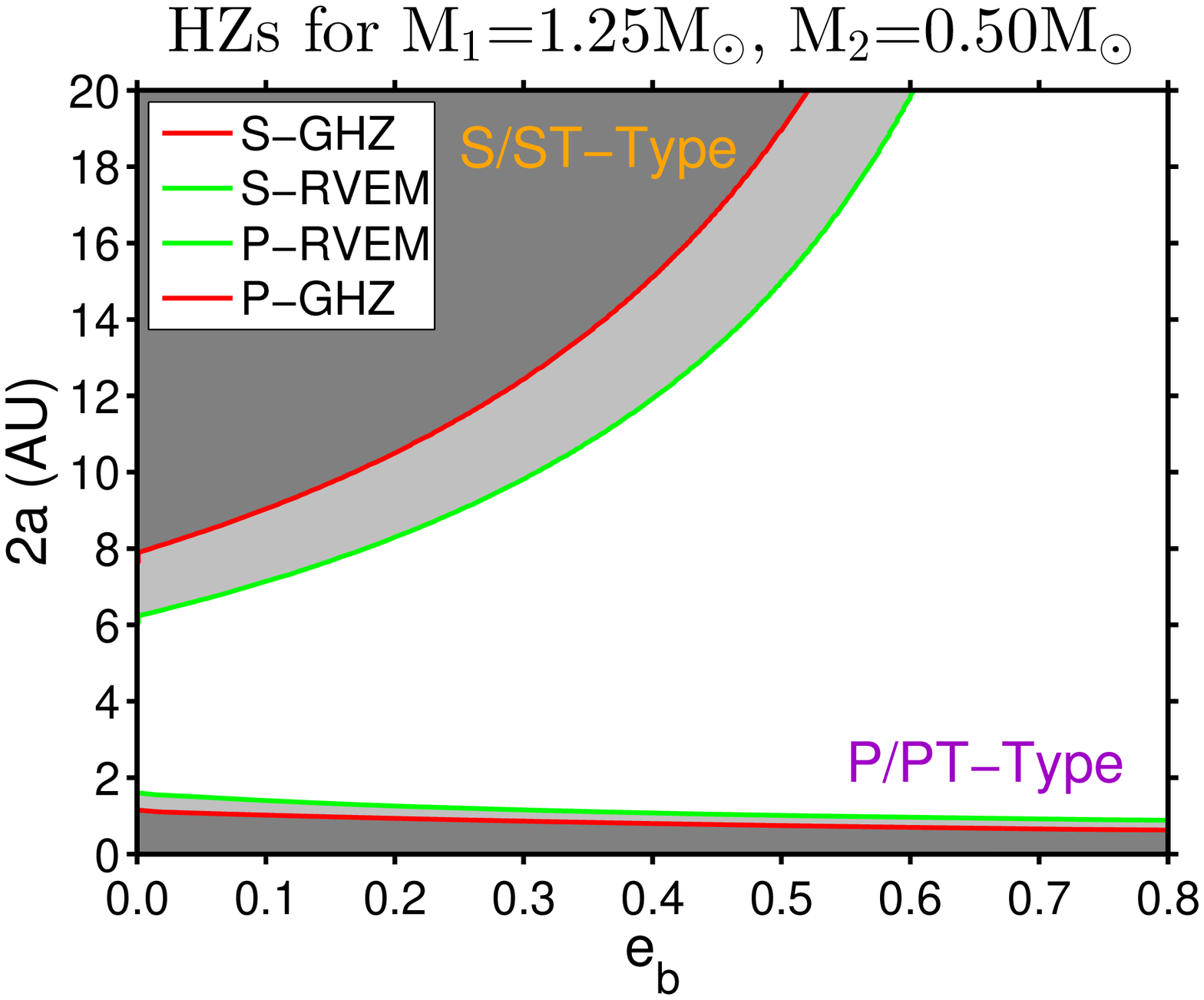} }}
	\caption{Required $2a$ and $e_{b}$ for GHZ and RVEM to exist regarding selected
binary systems. The GHZ can exist when the system parameters are within the gray region.
System parameters fall in either gray or light gray region would allow RVEM to exist.
The red and green curves show the critical pairs of values for the GHZ and RVEM to exist
correspondingly.}
	\label{fig:fig_narrow}
\end{figure*}

\begin{figure*}[t]
	\centering
	\subfloat{{\includegraphics[width=0.45\linewidth]{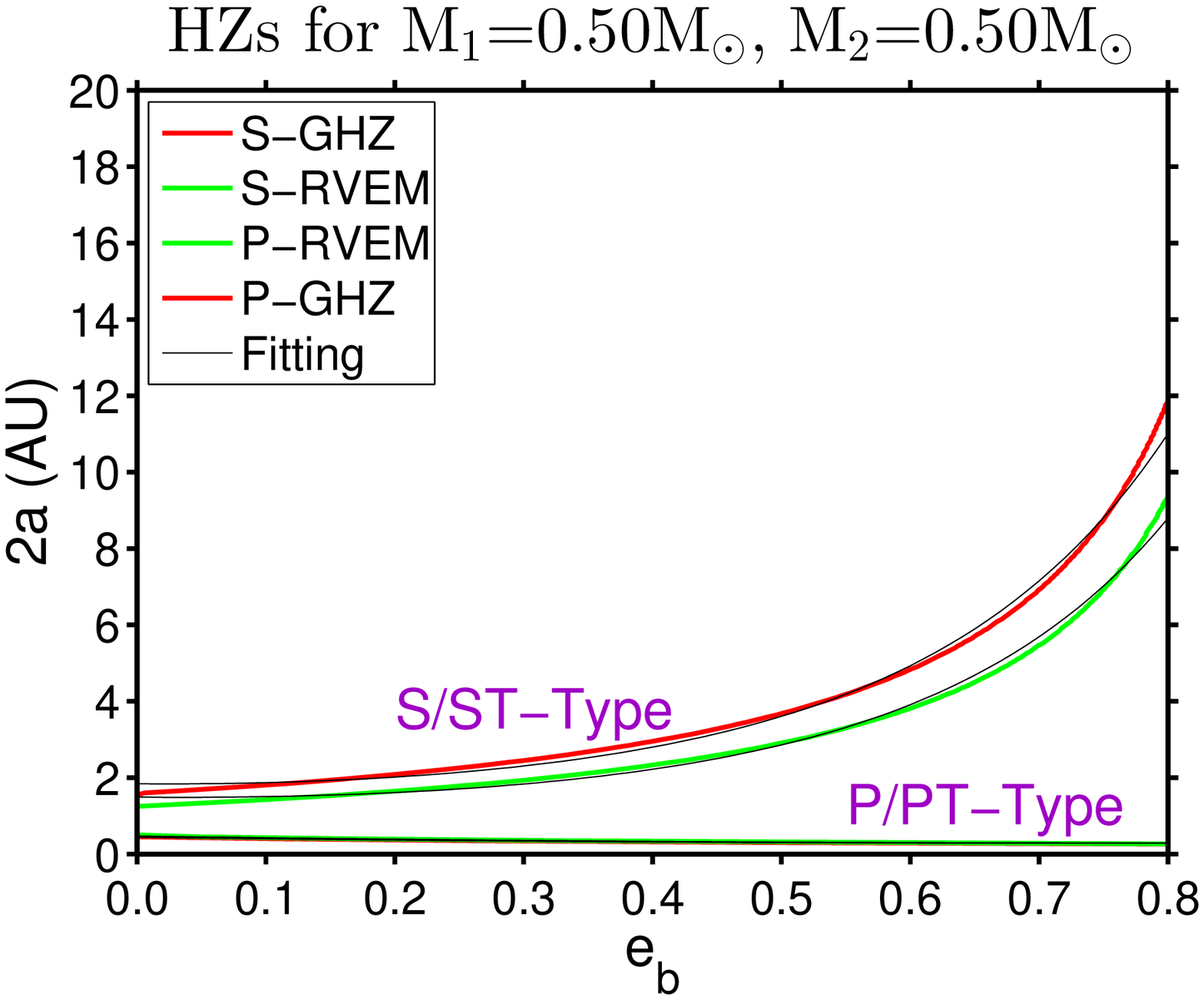} }}
	\subfloat{{\includegraphics[width=0.45\linewidth]{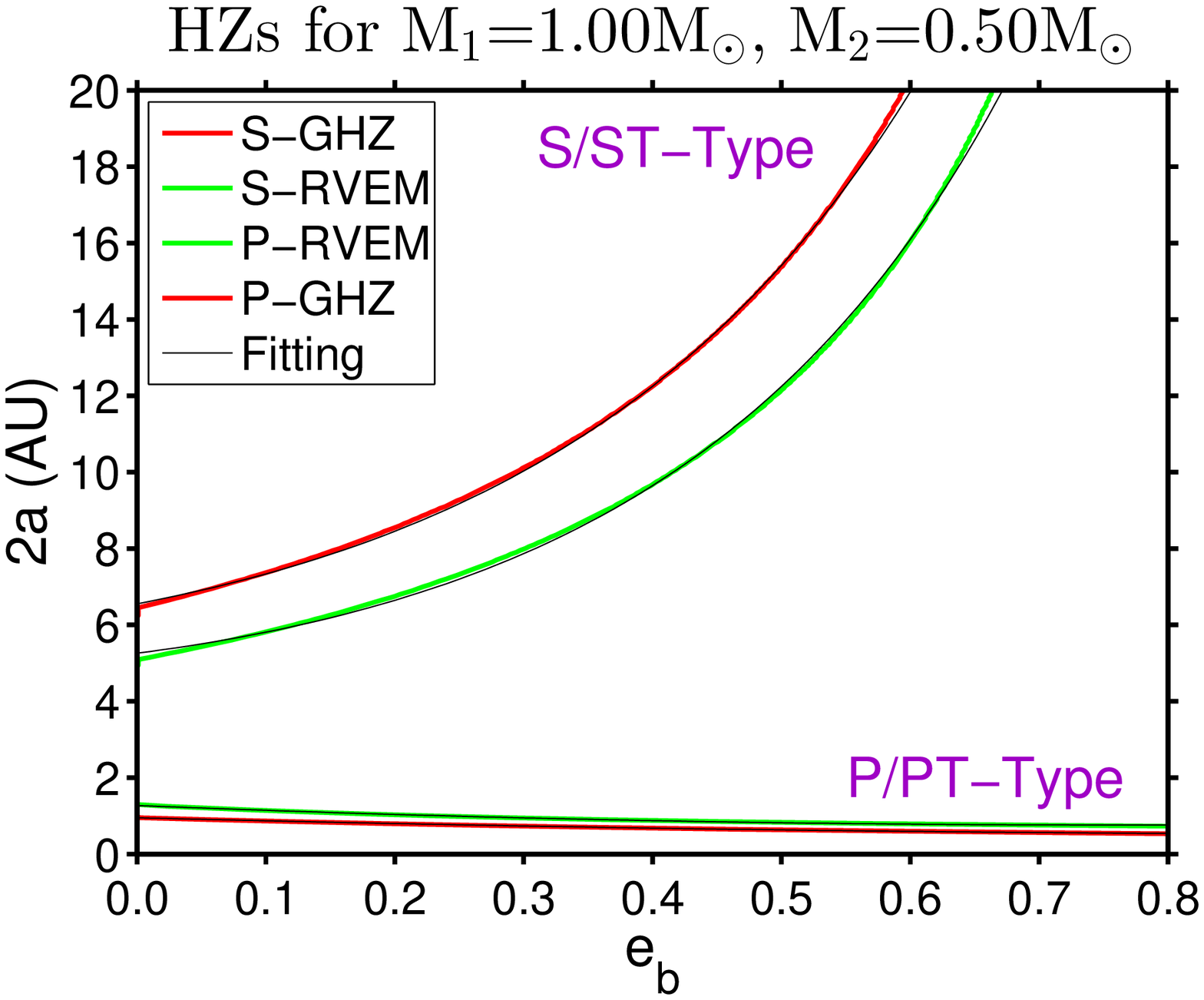} }}\\
	\subfloat{{\includegraphics[width=0.45\linewidth]{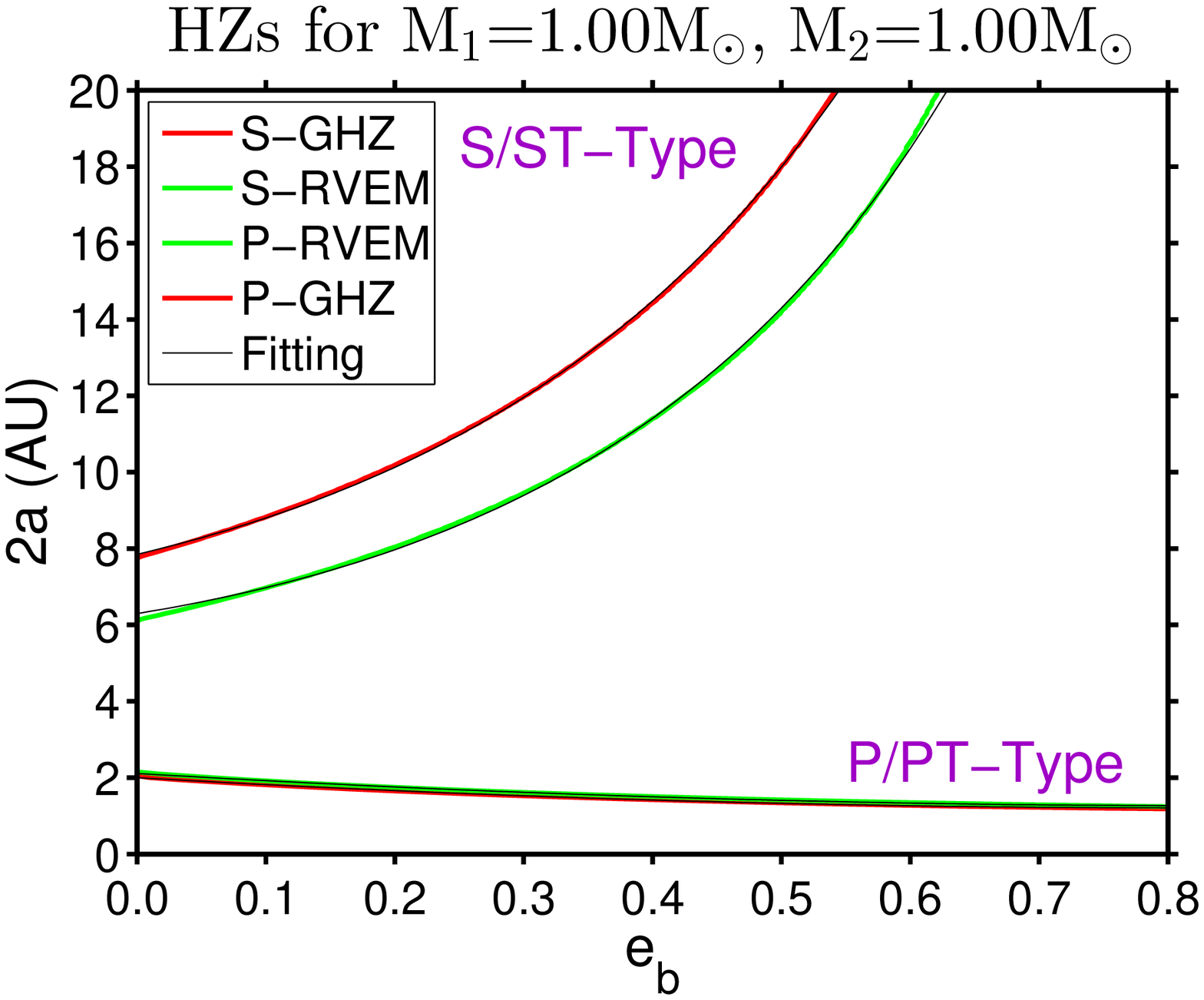} }}
	\subfloat{{\includegraphics[width=0.45\linewidth]{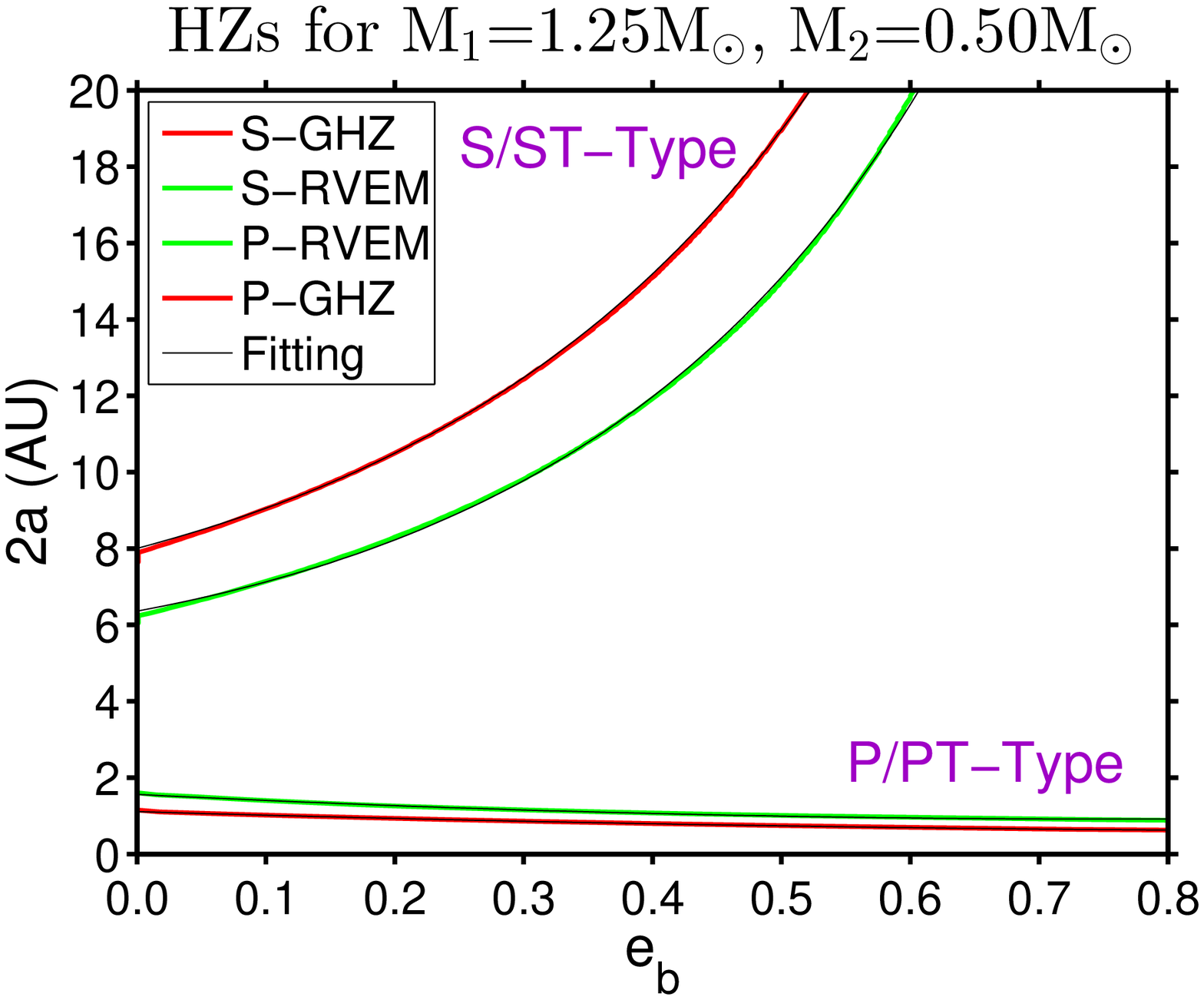} }}
	\caption{Fitting results compared with data for four cases.  The red and green lines
represent the boundaries for the GHZ and the RVEM to exist, respectively. In each subfigure,
the areas beyond the red and green curves (top) identify the existence of the S/ST-type HZs,
whereas the areas below the red and green curves (bottom) identify the existence of the
P/PT-type HZs. The thin black curves shown are the fitting results for the curve nearby,
and they are mostly indistinguishable from the data curves.}
	\label{fig:fig_narrow}
\end{figure*}

\section{Summary}
In this study, we explore the requirements for HZs to exist for selected examples
of binary systems based on the method given in Paper I and II with updated results
for terrestrial climate models obtained by Kopparapu et al. (2013, 2014).  Thus, we
developed fitting equations to efficiently determine the existence of HZs.  Utilizing
the fitting equations allows us to identify if the respective HZs is able to exist
without the need for cumbersome calculations.
Future work will deal with improving the fitting equations for enhanced accuracy.
We also plan to have $M_{1}$ and $M_{2}$ as parameters in the fitting
equations instead of being fixed values as for now.  This will make
the fitting equation more useful and applicable.

\section*{Acknowledgments}
{This work has been supported by the Department of Physics, University
of Texas at Arlington (UTA).}

\bibliographystyle{cs19proc}


\end{document}